\documentclass[letterpaper, 10 pt, conference]{ieeeconf}  % Comment this line out if you need a4paper

\IEEEoverridecommandlockouts            
\usepackage{graphicx} % for pdf, bitmapped graphics files
\usepackage{amsmath} % assumes amsmath package installed
\usepackage{subfigure}
\usepackage{booktabs}
\usepackage{multirow}
\usepackage[colorlinks,linkcolor=blue]{hyperref}

\title{\LARGE \bf
A Marker-free Head Tracker Using Vision-based Head Pose Estimation with Adaptive Kalman Filter
}

\author{
      Zhongxu Hu, \IEEEmembership{Member,~IEEE,},
      Chen Lv$^{*}$,\IEEEmembership{Senior Member,~IEEE,},
      Yanxin Zhou, Yiran Zhang,\\ and
      Wenhui Huang%
\thanks{$^{1}$Z. Hu, C. Lv, Y. Zhou, Y. Zhang and W. Huang are with the School of Mechanical and Aerospace Engineering, Nanyang Technological University, Singapore. (e-mail: {zhongxu.hu, lyuchen, yiran.zhang, yzhou031, huang.wenhui}@ntu.edu.sg)}%
}

\begin{document}

\maketitle
\thispagestyle{empty}
\pagestyle{empty}

%%%%%%%%%%%%%%%%%%%%%%%%%%%%%%%%%%%%%%%%%%%%%%%%%%%%%%%%%%%%%%%%%%%%%%%%%%%%%%%%
\begin{abstract}

The immersion and the interaction are the important features of the driving simulator. To improve these characteristics, this paper proposes a low-cost and mark-less driver head tracking framework based on the head pose estimation model, which makes the view of the simulator can automatically align with the driver's head pose. The proposed method only uses the RGB camera without the other hardware or marker. To handle the error of the head pose estimation model, this paper proposes an adaptive Kalman Filter. By analyzing the error distribution of the estimation model and user experience, the proposed Kalman Filter includes the adaptive observation noise coefficient and loop closure module, which can adaptive moderate the smoothness of the curve and keep the curve stable near the initial position. The experiments show that the proposed method is feasible, and it can be used with different head pose estimation models.

\end{abstract}

%%%%%%%%%%%%%%%%%%%%%%%%%%%%%%%%%%%%%%%%%%%%%%%%%%%%%%%%%%%%%%%%%%%%%%%%%%%%%%%%
\section{INTRODUCTION}

The intelligent driving is a currently hot research and trend, which requires a combination of multiple disciplines and multiple algorithms. Developing and testing algorithm in the real intelligent vehicles is an expensive and time consuming process\cite{ref_airsim}. The development of simulation technology provides a alternative way, the simulator can offer physically and visually realistic simulation for many research goals, it can also collect a large amount of annotated samples to leverage the deep learning and machine learning\cite{ref_ds1}\cite{ref_ds2}.

\par
\begin{figure}
    \centering
    \includegraphics[scale=0.23]{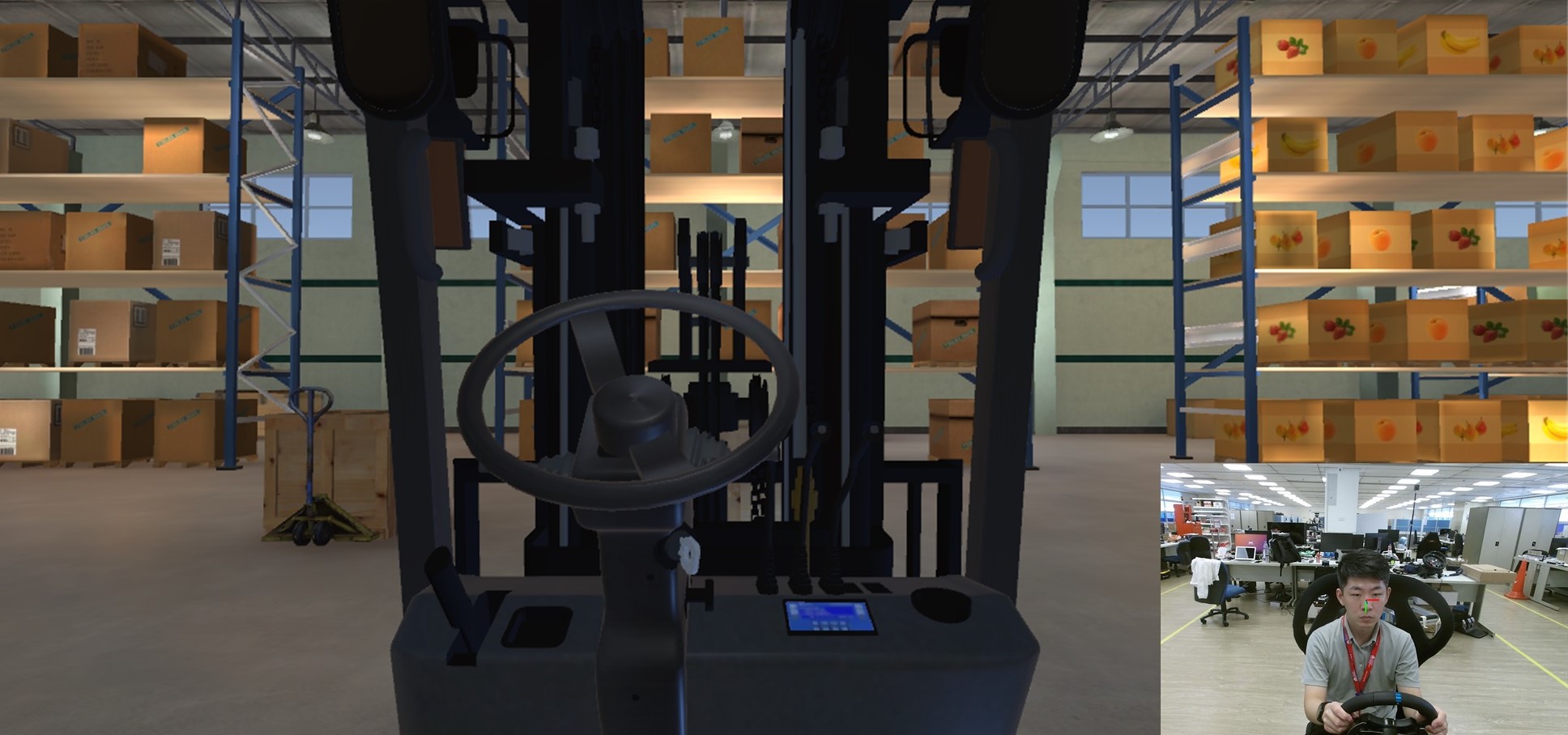}
    \caption{The software of the simulator.  It is used to simulate the forklift in a warehouse. The demo video could be found in the \href{https://www.youtube.com/watch?v=Kwdh2aUHsHU&feature=youtu.be}{Youtube website}.}
    \label{fig:sim}
\end{figure}

 The driving simulator cockpit is a widely used experimental platform. The one of the key features is the immersion. To improve the visual realism, a multi-screen mode is often used. But this will cause the distortion of the graphic ratio, and the field of the virtual view is limited and fixed. The cost of the multiple screens is also higher. Another way is to use virtual reality (VR) devices, this will bring two problems: 1. The dizziness caused by the serious mismatch between the fixed seat and the dynamic virtual graphic; 2. The VR glasses will cover the driver's face, which makes it impossible to carry out the research on the driver's state\cite{ref_db}\cite{ref_dratten}. Therefore, this paper aims to propose a low-cost adaptive view simulator solution through head tracking, as shown as Fig.\ref{fig:sim}.

\par The head pose as an important cue has been used in many human machine interaction fields. Nigel Sim et.al. proposed a wearable head-tracking device with inertial sensors as indicator of human movement intentions for Brain-Machine Interface application\cite{ref_headmouse}.  Chang Ho Kang et.al. presented a sensor fusion method which integrate the IMU, IR LED, CCD camera and other sensors\cite{ref_hybridtrack}. Adrian et.al. developed a low-cost head tracking device based on the SteamVR Tracking technology for virtual reality system\cite{ref_lighthouse}. These methods usually adopt different types of sensors to build the system. There are also several similar products in the flight simulators, such as TrackIR and Opentrack. They usually need some special device or optical marker, like IR Camera. Although the  newer FaceTrackNoIR only needs the RGB camera, they all require the user to manually adjust the relative parameters, and they usually use some traditional head estimation methods. This paper aims to adopt the state-of-the-art estimation model based on the deep learning which can improve the accuracy of the system.

\par To achieve the low-cost and mark-less solution, only the RGB sensor is used as the input device, and the head pose estimation model based on deep learning is adopted to improve the basic accuracy of the solution \cite{ref_headpose}. Due to the error of the model, the estimation curve of consecutive frames fluctuate sharply. The Kalman Filter is used to handle this problem \cite{ref_kf}. By analyzing the error distribution of the estimation model, an adaptive Kalman Filter is proposed to improve the performance of the solution, which includes an adaptive observation noise coefficient and a loop closure module, it can adaptive moderate the smoothness and keep the curve stable near the initial position. Finally, the proposed framework is verified by the designed experimental platform.

\par The main contribution of this paper as follows: 1. A simple, low-cost and effective framework of head tracking is proposed, which only uses the normal RGB camera; 2. According to the characteristics of the head pose estimation model, the adaptive Kalman Filter is proposed to improve the performance of the method.

\begin{figure*}[htb]
\centering
\includegraphics[scale=0.26]{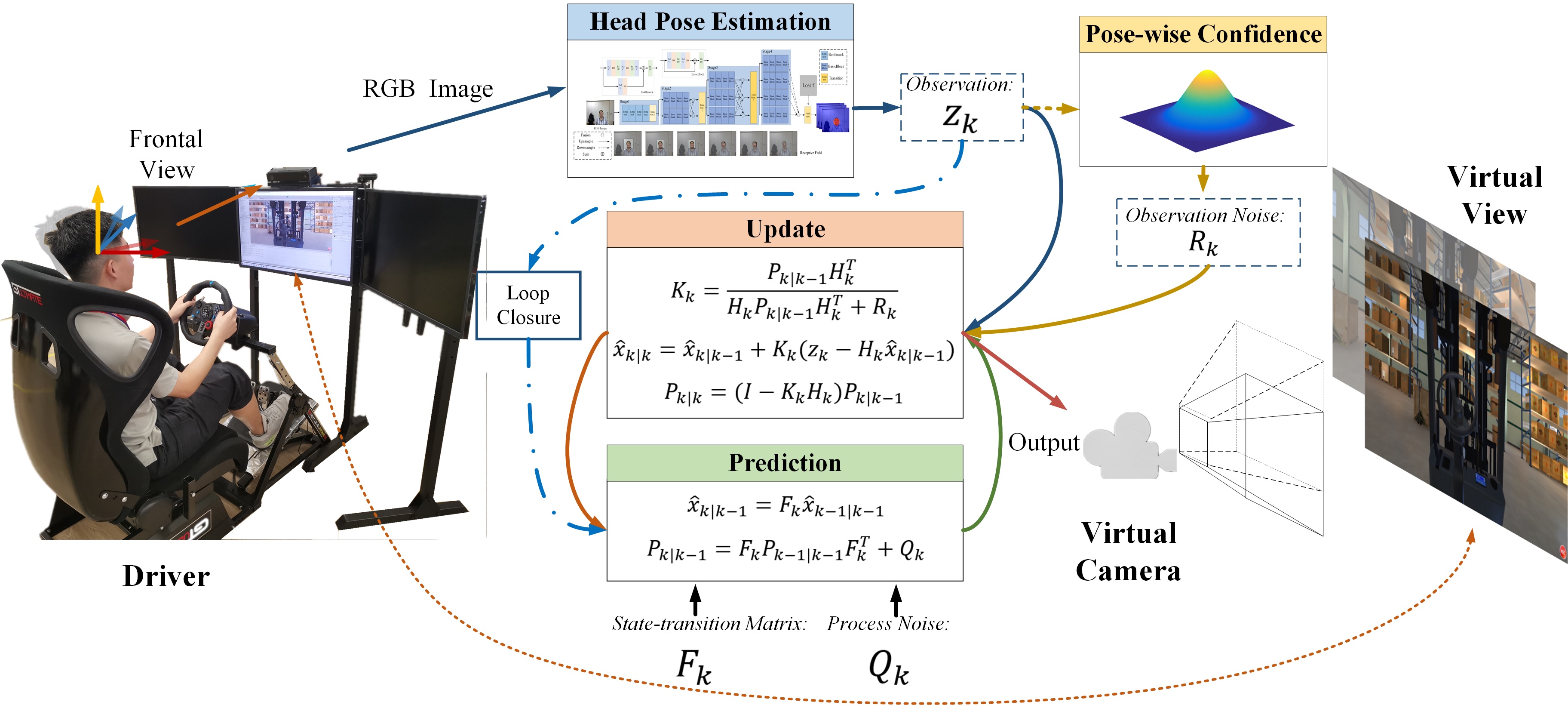}
\caption{The framework of proposed method. The left is the used driving cockpit which includes the input devices, computing server and RGB camera. The head pose estimation model is adopted as the measure, and its result as the observation. The proposed adaptive Kalman Filter is used to optimize the estimation. Finally, the virtual camera of the simulator is aligned with the output of the framework.}
\label{fig:Ov}
\end{figure*}

\par The paper is organized as follows. Section 2 describes the proposed head pose tracking method. Section 3 illustrates the designed  experiment platform and multiple experiment results. The conclusion and future work are in the Section 4.

\section{METHODOLOGY}

\subsection{Overview}
The purpose of this paper is to build a simulator whose view can be automatically adjusted with the driver's head pose based on frontal camera. The benefits are as follows: 1. It can improve the immersion and interaction of the simulator. The driver's view will be unconstrained and non-fixed, and the virtual camera will be synchronized with the driver's head pose; 2. The extracted head pose can also be used to monitor the state of driver; 3. It is a low-cost solution through a non-invasive camera sensor.
\par The development of deep learning and computer vision technology provides the basis for the proposed method. The state-of-the-art head pose estimation methods can achieve an error of $4-5^{\circ}$ on some public datasets. Although these methods work well, they cannot be directly applied to the simulator due to the inconsistency and volatility of the estimation. To solve this problem, this paper proposes a framework combining adaptive Kalman Filter (KF) and head pose estimation as shown as in Fig.\ref{fig:Ov}.
\par Assume that the true vector the driver's head pose at $k$ time is $\bar{x}_k=(\bar{p},\bar{y},\bar{r},\bar{v_p},\bar{v_y},\bar{v_r})$. The proposed method takes the estimation of the head pose estimation method based on deep learning as the observation vector $z_k=(\tilde{p},\tilde{y},\tilde{r})$. The posterior state estimation $\hat{x}_{k|k}=(\hat{p},\hat{y},\hat{r},\hat{v_p},\hat{v_y},\hat{v_r})$ is used as the coordinates of the virtual camera. To optimize the filter algorithm, this paper adopts an adaptive observation noise $R_k$ based on the different performance of the estimation model in different pose intervals. In addition, the proposed method also uses loop closure to make the virtual camera return to the initial position more effectively.

\subsection{Head pose estimation}

\begin{figure}[hbt]
    \centering
    \includegraphics[scale=0.18]{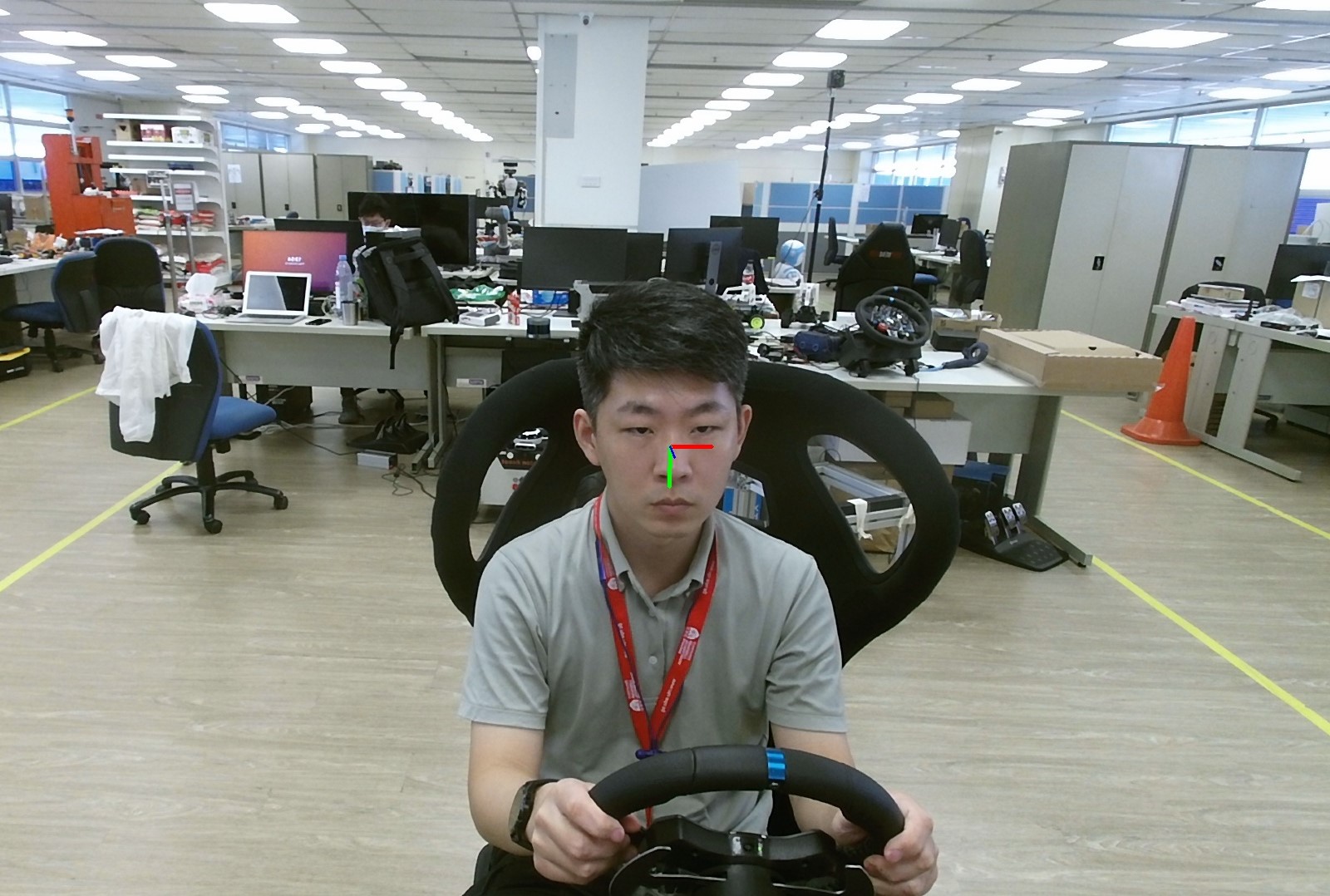}
    \caption{Head pose estimation. It includes $Yaw$,$Pitch$ and $Roll$.}
    \label{fig:head}
\end{figure}

Estimating the head pose is a crucial problem that has a large amount of applications, which is a task that needs to infer the 3D pose $(Pitch, Yaw, Roll)$ of the head from the input image. There are several different methods which use different input data, including depth image , RGB image and video clips. Considering the cost of hardware and computing, this paper will focus on the model based on a single RGB image.
\par With the development of the deep learning, the research on the head pose estimation has also achieved good results. In this paper, the different state-of-the-art head pose estimation models will be used to observe the driver's head. These methods have different performance and accuracy on different datasets, like BIWI\cite{ref_biwi}, AFLW2000\cite{ref_aflw2000} and AFLW \cite{ref_aflw} etc. The error range is about $3-5^{\circ}$. Using the different models are also to test the robustness of the proposed method.

\subsection{Adaptive Kalman filter}

\begin{figure}[hbt]
    \centering
    \includegraphics[scale=0.3]{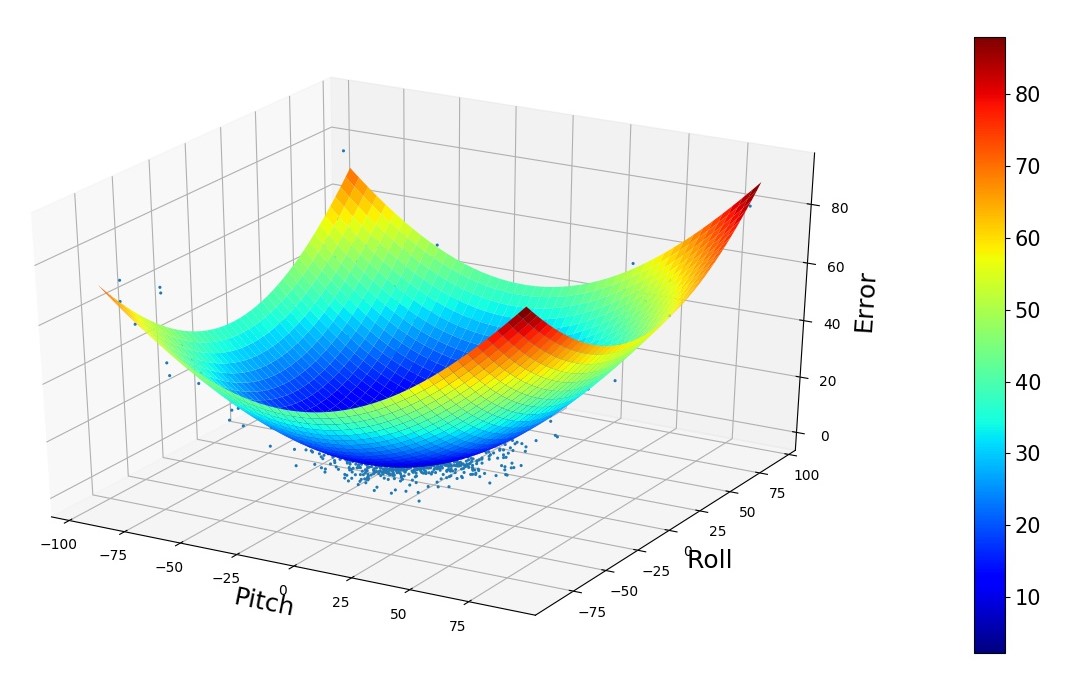}
    \caption{The error distribution of different intervals of $Pitch$ and $Roll$ based on AFLW2000 dataset. The 3d blue point represents the sample, and the curve surface is the results of 3d Gaussian fitting. }
    \label{fig:aflw_yr}
\end{figure}

As mentioned above, although the current head pose estimation method has good performance, there is still a certain error. When it is applied to the simulator, its flaws of fluctuation and discontinuity will be highlighted. From a practical perspective, the smoothness and continuity of the view changes are more important than the accuracy. To solve this problem, the kalman filter is adopted. Kalman filtering is an algorithm that provides estimates of some unknown variables, that tend to be more accurate, given the measurements observed over time and contained statistical noise and other inaccuracies. Kalman filters have been demonstrated its usefulness in various applications, such as guidance, navigation, and control of vehicles etc. Kalman filters have relatively simple form and require small computational power.
\par To use the KF, the problem needs to be modeled, and it is assumed as a linear model to ensure real-time performance. Assuming that the posterior state estimation of the head pose at $k$ time is $\hat{x}_{k|k}=(\hat{p},\hat{y},\hat{r},\hat{v_p},\hat{v_y},\hat{v_r})$. The prediction phase is as follows: 

\begin{gather}
    \hat{x}_{k|k-1}=F_k\hat{x}_{k-1|k-1} \\
    P_{k|k-1}=F_kP_{k-1|k-1}F^T_k+Q_k \\
F_k=\begin{bmatrix}
I_{3\times 3} & \Delta t \cdot I_{3\times 3}\\ 
0 & I_{3\times 3}
\end{bmatrix}
\end{gather}

where the $\hat{x}_{k|k-1}$ is the prior state estimation, $F_k$ is the state-transition matrix,$P_{k|k-1}$ is the prior state estimation covariance at $k$ time, while the $P_{k-1|k-1}$ is the posterior estimation covariance at $k-1$ time.$Q_k$ means the process noise. The $I_{3\times 3}$ represents the identity matrix of $3\times 3$ size.Then the update phase is:

\begin{gather}
    K_k=\frac{P_{k|k-1}H^T_k}{H_kP_{k|k-1}H^T_k+R_k} \\
    \hat{x}_{k|k}=\hat{x}_{k|k-1}+K_k(z_k-H_k\hat{x}_{k|k-1}) \\
    P_{k|k}=(I-K_kH_k)P_{k|k-1} \\
    H_k=\begin{bmatrix}
I_{3\times 3} & 0 \\ 
\end{bmatrix}
\end{gather}

The $K_k$ is the Kalman gain factor, $H_k$ means the measurement matrix which converts the state variable into the corresponding observation variable. The $\hat{x}_{k|k}$ is the posterior state estimation, which  is also the pose of the virtual camera of the simulator. The $z_k$ is the output of the head pose estimation model, which as  the observation value. The $P_{k|k}$ is the posterior estimation covariance at $k$ time. The $R_k$ is the observation noise covariance, which is related 
to the  estimation model and will also affect the performance of the filter.

\par To determine the $R_k$, the several head pose estimation models are analyzed. Through statistics, it is found that these models have different performances in different intervals of the head pose.Usually, the accuracy is higher when the pose angle is small, otherwise the error is higher, especially $Yaw$ and $Roll$. For example, the AFLW2000 dataset, a widely used benchmark, is used to test the head pose estimation model, and the result is as shown as in the Fig.\ref{fig:aflw_yr}. The $Yaw$ and $Roll$ of the samples are taken as the $X$ and $Y$ axis, and the $Error$ is taken as the $Z$ axis. The blue 3d points represent the different samples. A 2d Gaussian function is used to fit the points as shown as the curved surface in the Fig.\ref{fig:aflw_yr}.

\par Therefore, the adaptive $R_k$ is proposed by this paper as follows: 

\begin{equation}
R_k(x)=\tau - \lambda \cdot \frac{1}{\sqrt{2\pi }\sigma }exp(-\frac{(x-\mu )^2}{2\sigma ^2})
\label{eq_rk}
\end{equation}

Where the $\tau$ means the offset, the $\lambda$ represents the amplitude factor, the $\sigma$ and $\mu$ are the mean and variance respectively. So the $R_k$ can be adaptively adjusted in the iterative process. It can make the filtered value close to the observed value when the rotation angle is small, while the filtered value changes smoother when the rotation angle is large.

\begin{figure*}[htb]
    \centering
     \subfigure[FSA-Net]{
        \label{aflw_fsa}
        \includegraphics[scale=0.4]{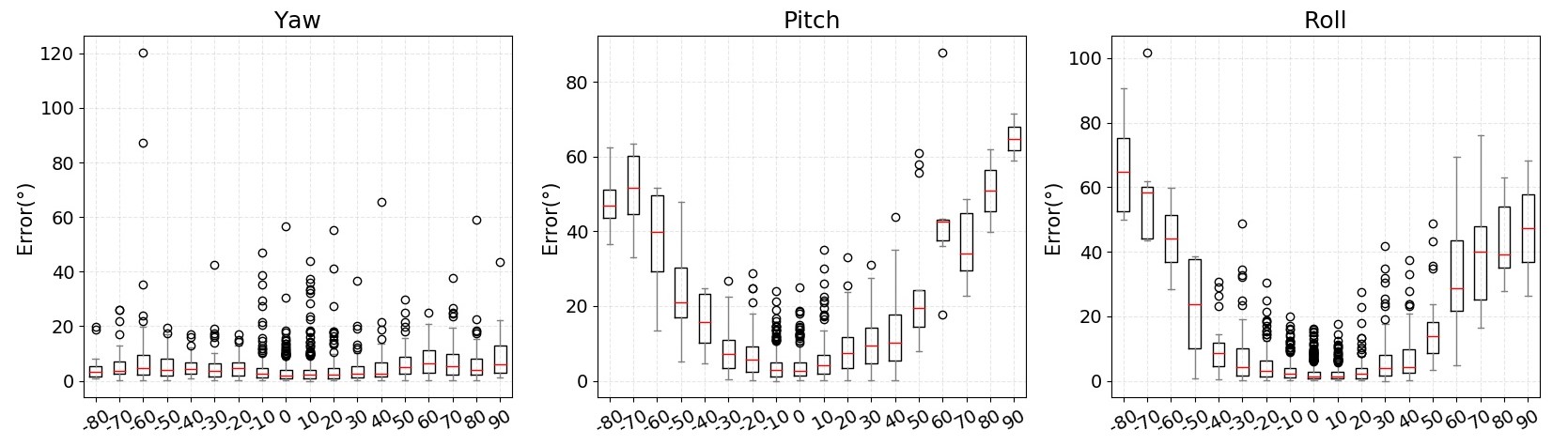}}
    \subfigure[Hopenet]{
        \label{aflw_ssr}
        \includegraphics[scale=0.4]{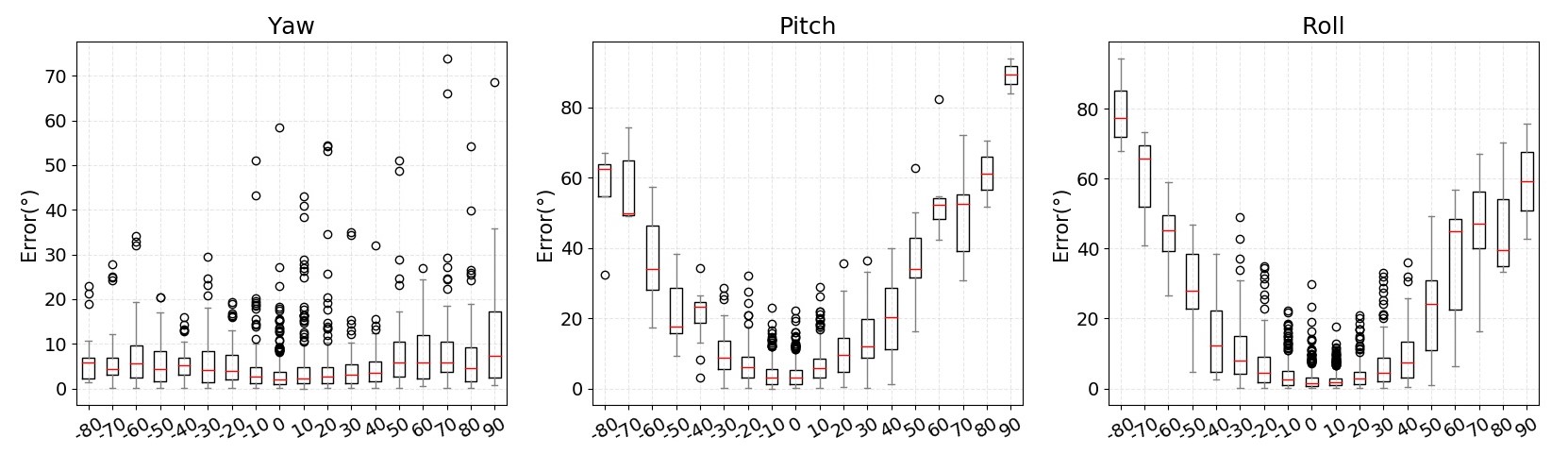}}
    \caption{The error distributions of different model in the AFLW2000 dataset. The X axis means the angle intervals, the Y axis represents the corresponding error.}
    \label{fig:aflw}
\end{figure*}

\subsection{Loop closure}
There are two important points for the experience of use: First, the rotation of the simulator virtual camera is smooth. The smoothness of the filter can be changed by adjusting the $Q_k$. Based on this, the adaptive $R_k$ makes the filter close to the real value. Another point is whether the virtual camera can accurately return to its initial position in time. The driver spends more time in a relatively position, such as looking ahead. Moreover, the driver is more sensitive to whether the virtual camera is back-aligned than whether it is rotated to the correct angle. To handle this situation, the loop closure is added into the proposed pipeline.

\par The loop closure is widely used in the Simultaneous localization and mapping (SLAM) algorithm. Usually, the purpose is to reduce the accumulative error by detecting whether the agent has return to a previously visited position. In this paper, the loop closure can be used to maintain the virtual camera close to the initial position by fusing the observed value with the initial value when they are  close.

\begin{equation}
   \label{eq_zk}
    z_k=\left\{\begin{matrix}
\xi \cdot  z_k+(1-\xi)\cdot \kappa, s.t.\left \| z_k-\kappa \right \| \leq \theta \\ 
z_k, s.t.\left \| z_k-\kappa \right \| >  \theta
\end{matrix}\right.
\end{equation}

where the $\kappa$ is the initial value, the $\xi$ is the fusion factor and the $\theta$ means the threshold. When the driver moves in a small range near the initial position, the simulator will keep the initial scene as soon as possible. This is more in line with actual experience.

\section{Experiments}

\subsection{Experimental platform}
The purpose of this paper is to improve the driving simulation. So a driving simulator is used as the experiment platform as shown as the left of the Fig.\ref{fig:Ov}. The simulator includes a computing server (RAM 32GB, CPU i7, GPU 2080), a set of input devices (Logitech G29), a RGB Camera (Kinect V2). The camera does not need to be calibrated.

\par The software of the simulator is developed by the Unity3D and Airsim, which can simulate the several operations of the forklift in the warehouse scenario as shown as the Fig.\ref{fig:sim}. So the virtual camera can be controlled easily. Finally, more than 10,000 driver frontal images have been collected like the Fig.\ref{fig:head}, and they are consecutive frames.

\subsection{Head pose estimation models}
To test the robustness of the proposed pipeline, two different head pose estimation models, the FSA-Net\cite{ref_fsa} and the Hopenet\cite{ref_hn}, are adopted in the next experiments, which have different performance in some open datasets. The FSA-Net is one of the state-of-the-art methods, which proposes a fine-grained structure mapping for spatially grouping features before aggregation. The Hopenet presents a robust way to determine pose by training a multi-loss model on a large synthetically expanded dataset to predict intrinsic angles (yaw, pitch and roll) directly from image intensities through joint binned pose classification and regression.

\begin{table}[!htbp]
\centering
\caption{Experiments to calculate the Gaussian fitting parameters of different models based on the AFLW2000 dataset}\label{tab_comparison}
\begin{tabular}{lccccc}
\toprule
\textbf{Methods} & \textbf{Angles} & \textbf{ $\lambda$ } & \textbf{$\mu$} & \textbf{$\sigma$} & \textbf{ $\tau$}\\
\midrule
\multirow{3}*{FSA-Net} & Yaw  & 4.11 & -0.35 & 30.87 & 7.64 \\
~ & Pitch & 312.07 & -5.19 & 132.41 & 315.43\\
~ & Roll & 3.29e+05 & -5.62e-01 & 4.44e+03 & 3.29e+05 \\
\midrule
\multirow{3}*{Hopenet} & Yaw & 7.017 & -5.57 & 48.28 & 10.74 \\
~ & Pitch & 229.18 & -8.30 & 101.37 &232.88 \\
~ & Roll & 9.35e+04 & 4.76e-02 & 2.219e+03 & 9.35e+04 \\
\bottomrule
\end{tabular}
% \begin{tablenotes}
% \footnotesize
% %\item[1] $\uparrow$ means that the larger value, the better performance.
% \end{tablenotes}
\end{table}

\par The AFLW2000 is a widely used benchmark, which contains 2000 different head pose images with different real-world backgrounds and light conditions. In this section, it is used to determine the error distribution of the head pose estimation model in different pose intervals. The results are as shown as the Tab.\ref{tab_comparison} and Fig.\ref{fig:aflw}. Overall, the FSA-Net performs better, which has a lower error in different angles. But what they have in common is that the Gaussian distribution is more obvious on the $Pitch$ and $Roll$. Then the Eq.\ref{eq_rk} can be used to fit these distributions.

\subsection{Comparison}

\begin{figure}[hbt]
    \centering
    \subfigure[Pitch]{
    \label{kf_p}
    \includegraphics[scale=0.23]{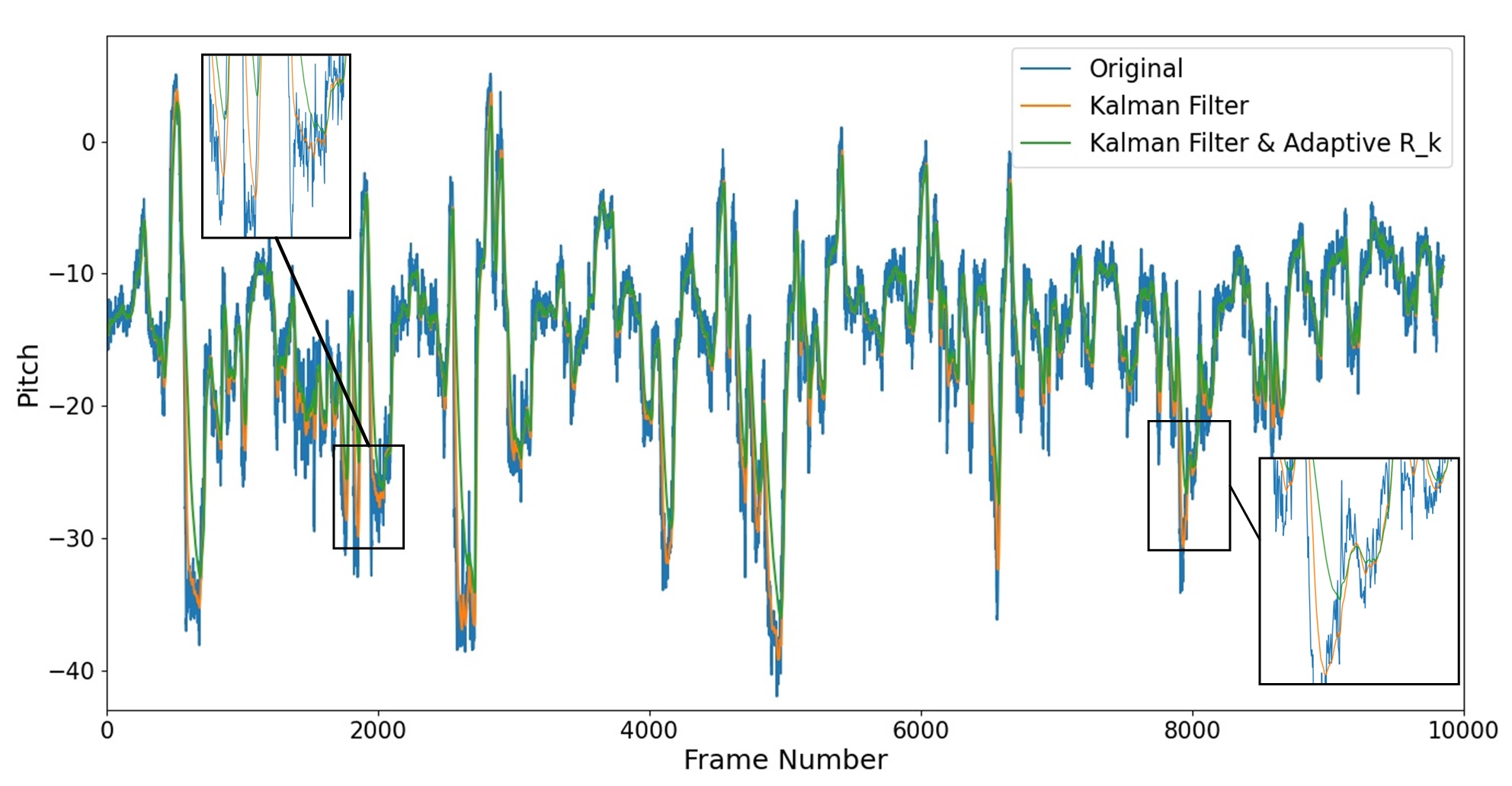}
    }
    \subfigure[Roll]{
    \label{kf_r}
        \includegraphics[scale=0.23]{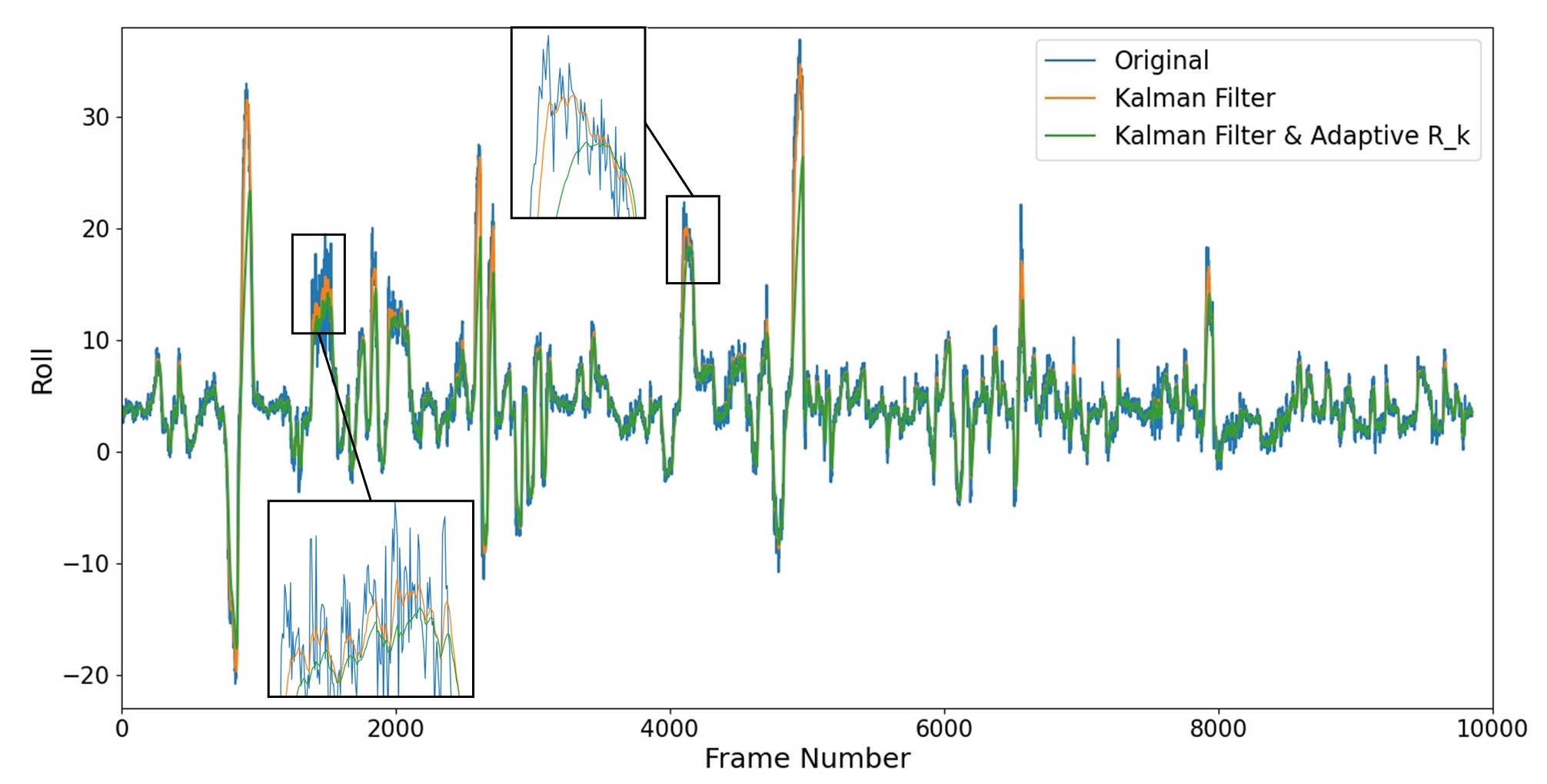}
    }
    \caption{The comparison of original data, standard Kalman Filter and the proposed Kalman Filter with adaptive $R_k$ in the $Pitch$ and $Roll$ angles domain.}
    \label{fig:kf_comp_yr}
\end{figure}

\begin{figure}
    \centering
    \includegraphics[scale=0.23]{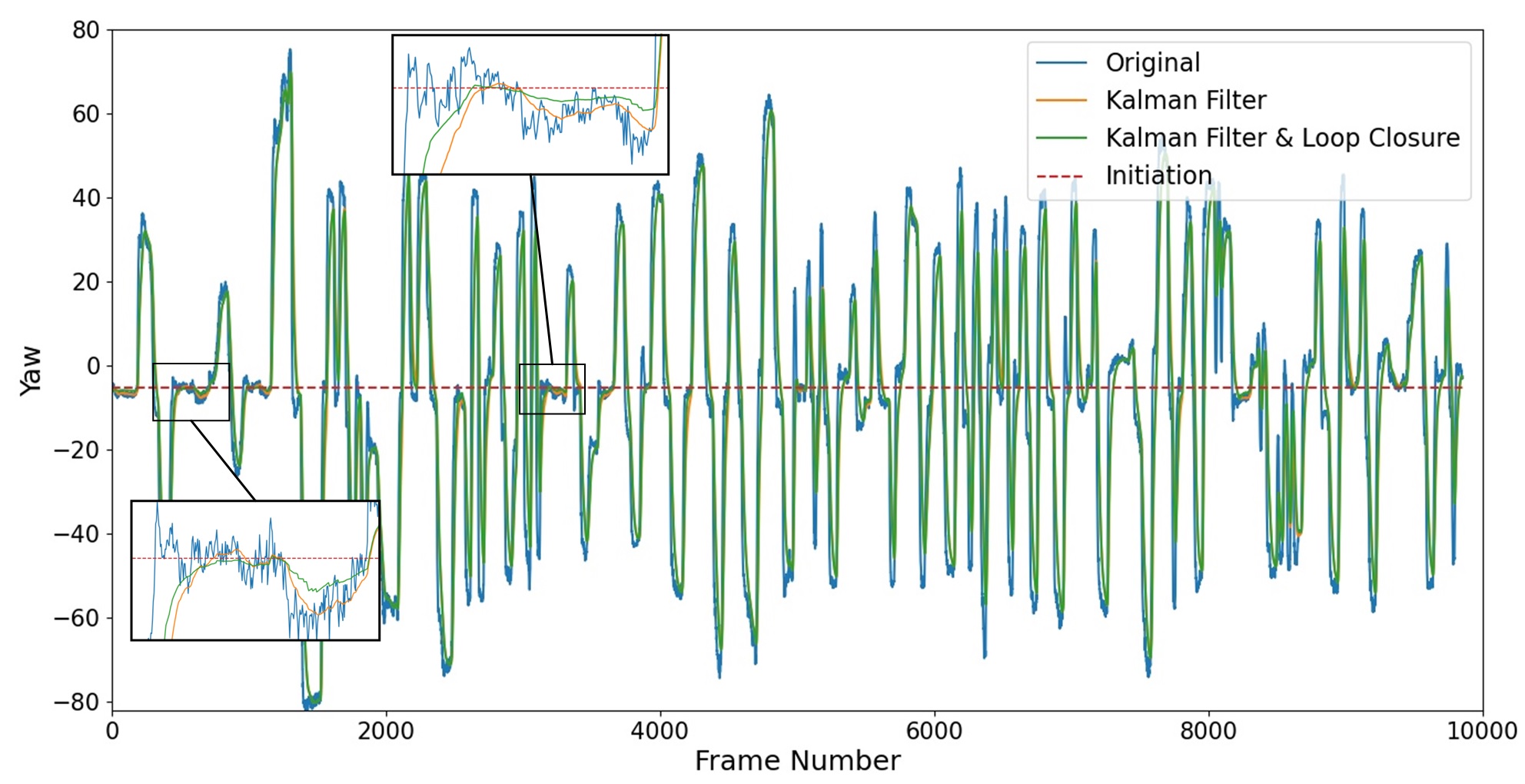}
    \caption{The comparison of original data, standard Kalman Filter and the proposed Kalman Filter with loop closure in the $Yaw$ angle domain}
    \label{fig:kf_comp_lc}
\end{figure}

\begin{figure}[hbt]
    \centering
    \subfigure[Yaw]{
    \label{kf_y_comp}
    \includegraphics[scale=0.23]{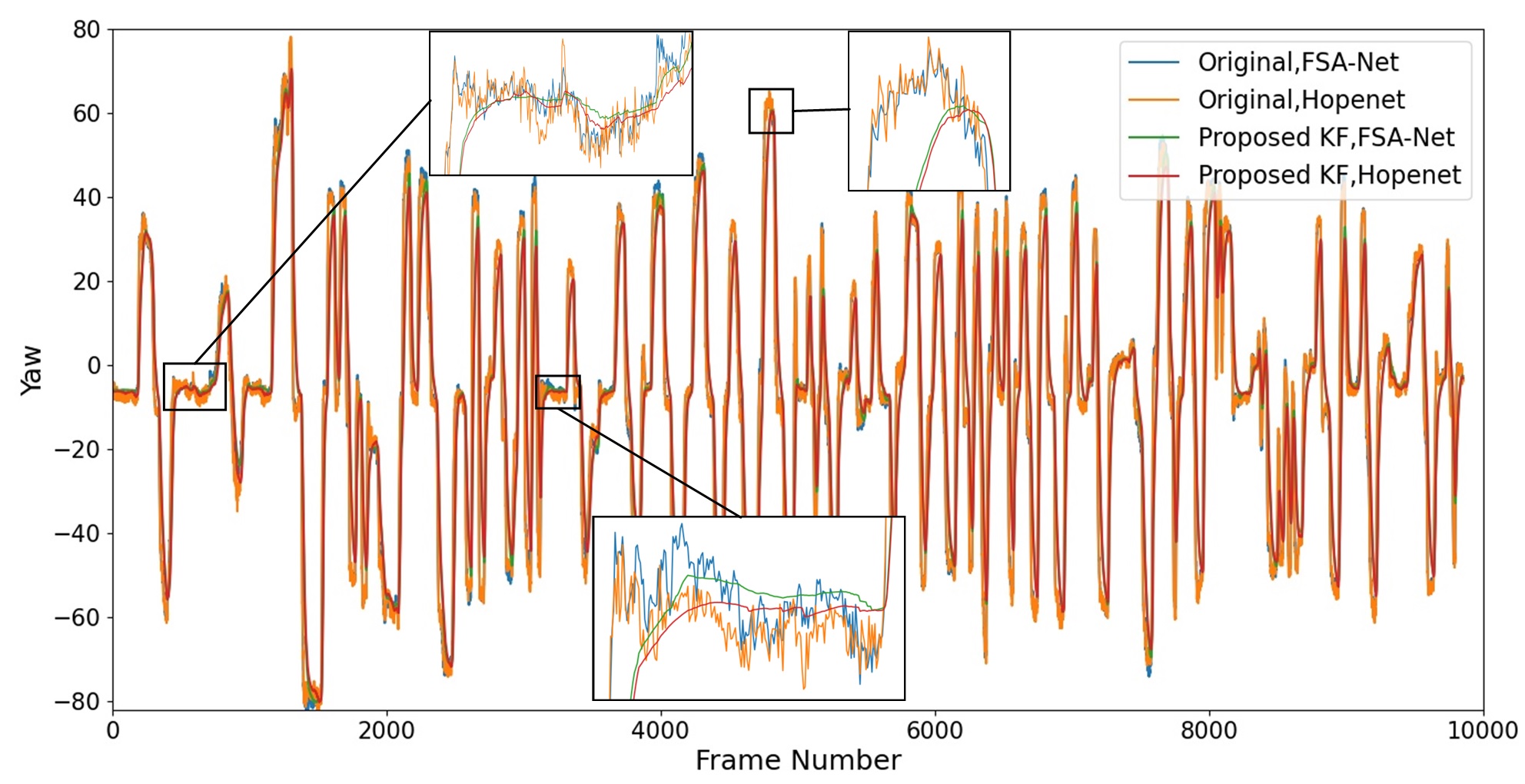}
    }
    \subfigure[Pitch]{
    \label{kf_p_comp}
        \includegraphics[scale=0.23]{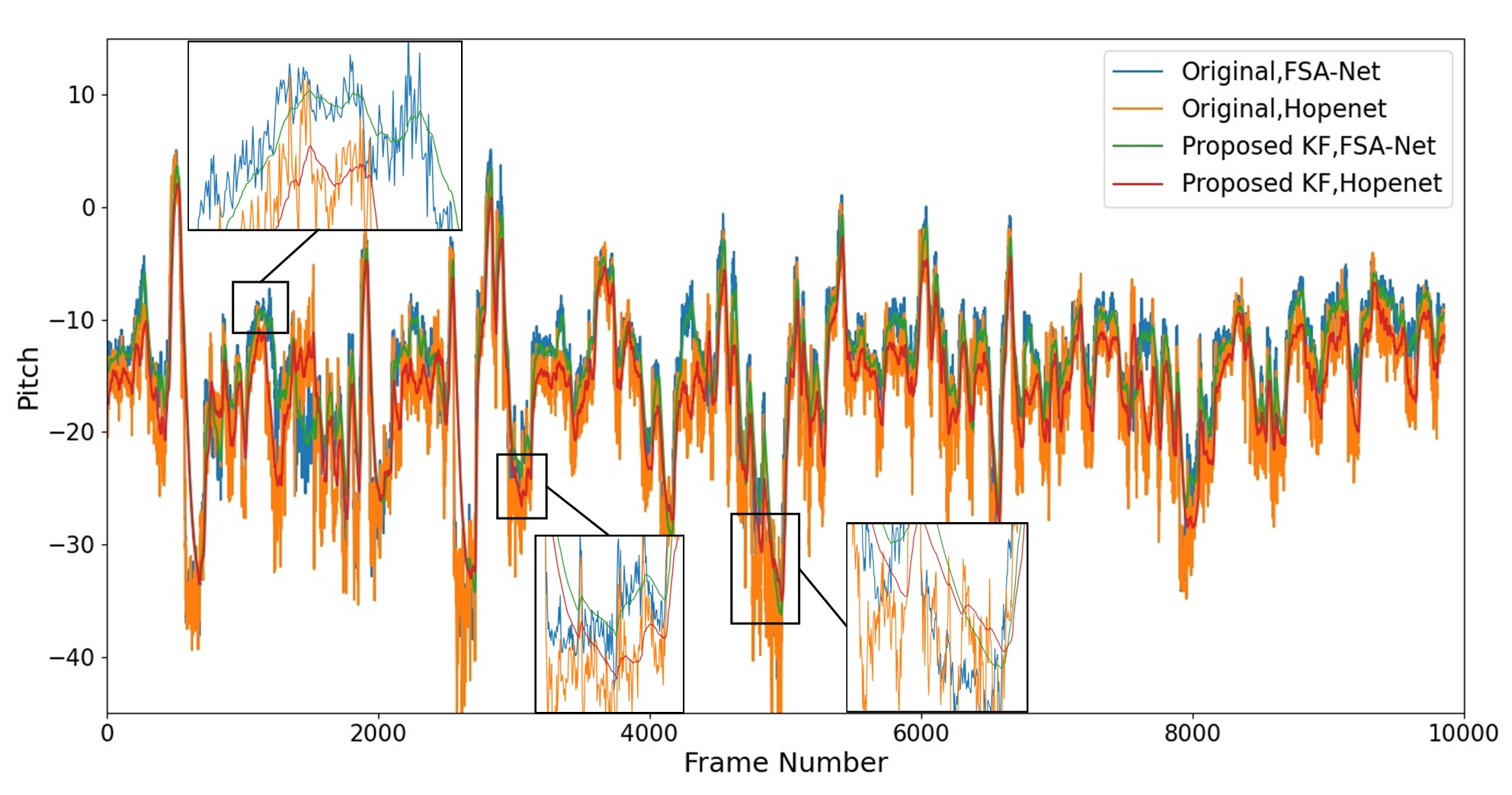}
    }
    \subfigure[Roll]{
    \label{kf_r_comp}
        \includegraphics[scale=0.23]{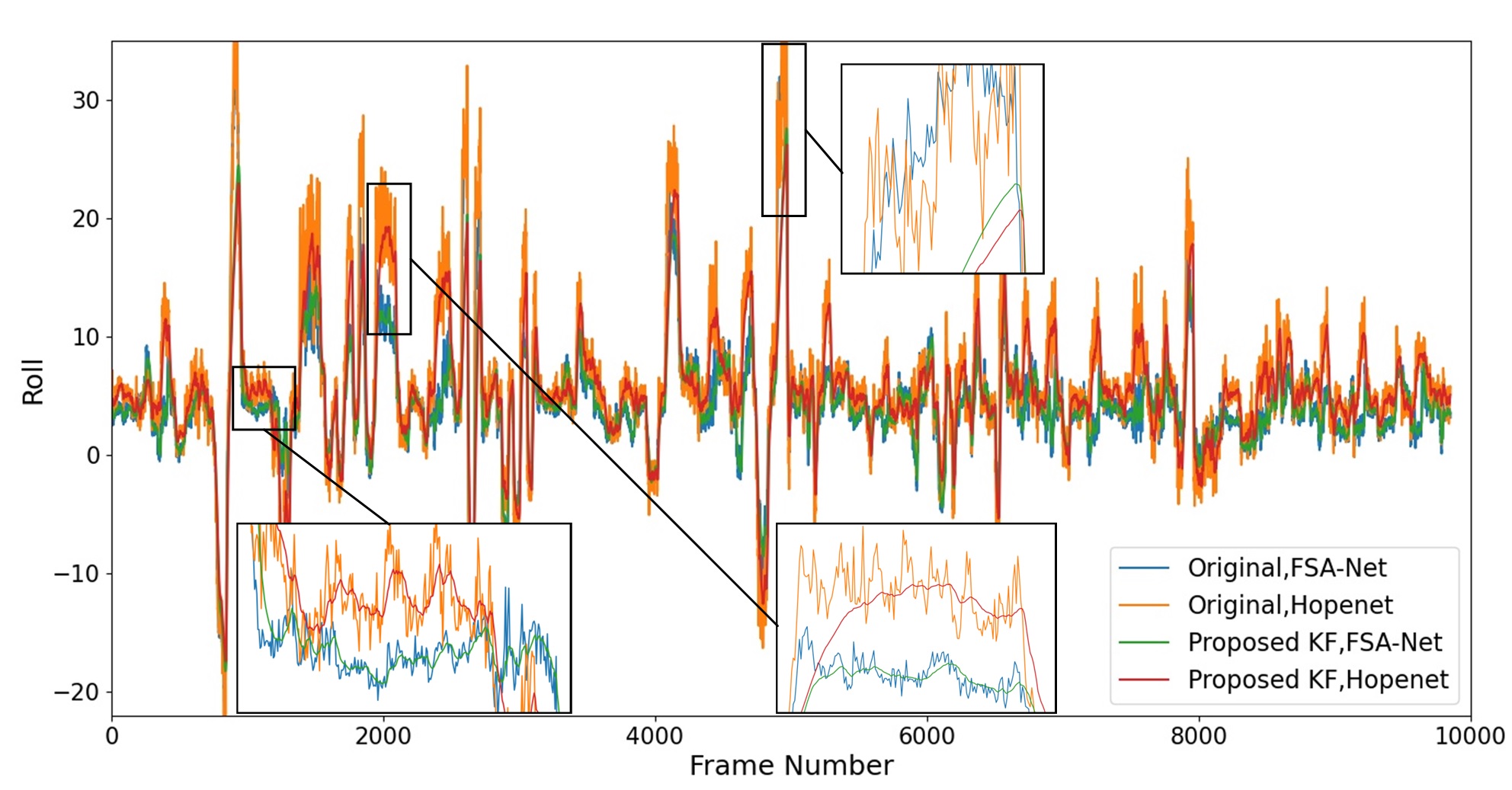}
    }
    \caption{The comparison of the proposed Kalman Filter with different head estimation models.}
    \label{fig:kf_comp}
\end{figure}

To evaluate the proposed pipeline, the FSA-Net is first used to estimate the driver head pose based on the  collected driver frames, the results are as shown as the \textit{Original} of the  Fig.\ref{fig:kf_comp_yr}. Obviously, these original data can not be directly used for the simulator. After the Kalman Filter is used, the curve becomes smooth and the volatility is significant reduced. This shows that it is necessary and reasonable to use the Kalman Filter. At this time, the $R_k$ is a constant value, which is the mean value of the Gaussian function that is calculated in the last subsection. For further improve the performance, the constant $R_k$ is replaced as the adaptive one by the mentioned in the Eq.\ref{eq_rk}, and the related parameters are the results of the Gaussian fitting on the AFLW2000 dataset. The comparisons are as shown as the Fig.\ref{fig:kf_comp_yr}. It can be seen that the curves of the standard Kalman Filter and the filter with adaptive $R_k$ are almost coincide at the low angle. But the different is that the filter with adaptive $R_k$ has better performance at high angle, the curve is more smooth. This is the advantage of the adaptive $R_k$. It is worth noting that only the curves of $Pitch$ and $Roll$ are listed in the Fig.\ref{fig:kf_comp_yr} for a more intuitive comparison.

\par Due to the driver often returns to the initial position, and it is most obvious in the $Yaw$ domain as shown as the Fig.\ref{fig:kf_comp_lc}. To evaluate the performance of the loop closure, the standard KF is modified by the Eq.\ref{eq_zk}, and the $\xi$ is 0.618 and the $\theta$ is 2. It can be seen that the loop closure can make the angle more stable and close to the initial value when the angle is near the initial value. In the 
remain angles, the curves of standard KF and the filter with loop closure are coincide.

\par The purpose of this paper is to propose a framework which takes the head pose estimation model as the input. The framework could handle different estimation models, which means that the proposed framework can be optimized as the head pose estimation technology improves. So the two different head pose estimation models are used to handle the same driver frames as shown as in the Fig.\ref{fig:kf_comp}. It can be seen that the original outputs of the models have large differences and fluctuations. The proposed method can significantly reduce this kind of deviation, the trends of the curves are basically same. It shown that the different head pose estimation model with different accuracy can be used in the proposed framework.

\section{Conclusion}
To improve the immersion and the interaction of the driving simulator, this paper proposes a framework which can make the view of the simulator automatically change with the driver's head pose. The proposed method only uses the RGB camera without the other hardware or marker. The challenge is that the currently head pose estimation methods still have certain errors, which can not be directly adopted. To handle this problem, this paper proposes a framework which combines the head pose estimation and the Kalman Filter. By analyzing the error distribution of the estimation model and user experience, the adaptive Kalman Filter is proposed, which includes the adaptive observation noise coefficient $R_k$ and loop closure module. The adaptive $R_k$ makes the curve smoother in the area  where the estimation model has large error, and the loop closure makes the curve stable near the initial position. The experiments show that the proposed method is feasible, and it can be used with different head pose estimation models. The demo also can be found in the Youtube website.

\par This paper proposed a reasonable framework and solution for head tracking, it can be further optimized as algorithm of head pose estimation improves. It also can be used in the other human-in-the-loop applications, and the source code of this paper will be open sourced.

\section*{ACKNOWLEDGMENT}

This work was supported by the A*STAR Grant (No. 1922500046), Singapore.

%%%%%%%%%%%%%%%%%%%%%%%%%%%%%%%%%%%%%%%%%%%%%%%%%%%%%%%%%%%%%%%%%%%%%%%%%%%%%%%%

\end{document}